\documentclass[notitlepage,nofootinbib]{revtex4}
\usepackage[utf8]{inputenc}

\usepackage{subcaption}
\usepackage{natbib}
\usepackage{graphicx}
\usepackage{epsfig}
\usepackage{amsmath}
\input{epsf}
\usepackage{psfrag}

\usepackage[usenames,dvipsnames]{color}

\newcommand{\beq}{\begin{eqnarray}}
\newcommand{\eeq}{\end{eqnarray}}
\newcommand{\Slash}[1]{{\ooalign{\hfil/\hfil\crcr$#1$}}}

\newcommand{\nn}{\nonumber \\}

\usepackage{amsmath,color}
\usepackage{amssymb}
\usepackage{bm}
\usepackage{graphicx}
\usepackage{multirow}

\begin{document}

\title{\boldmath
$\phi$-meson lepto-production near threshold and the strangeness $D$-term 
}


\author{Yoshitaka Hatta}

\affiliation{Physics Department, Brookhaven National Laboratory, Upton, New York 11973, USA}
\affiliation{ RIKEN BNL Research Center, Brookhaven National Laboratory, Upton, New York 11973, USA}

\author{Mark Strikman}
\affiliation{ Department of Physics, Penn State University, University Park PA 16802, USA}


\date{\today}

\begin{abstract}

We present a model of exclusive $\phi$-meson lepto-production $ep \to e'p'\phi$ near threshold  which features the strangeness  gravitational form factors of the proton. We argue that the shape of the differential cross section $d\sigma/dt$ is a sensitive probe of the strangeness D-term of the proton.

\end{abstract}

\maketitle

\section{Introduction}

Exclusive lepto-production of vector mesons $e p\to e'\gamma^* p \to e'p'V$ is a versatile process that can address a wide spectrum of key questions about the proton structure. Depending on the $\gamma^*p$ 
 center-of-mass energy $W$, photon virtuality $Q^2$ and meson species, the spacetime picture of the reaction looks  different and  requires different  theoretical frameworks. At high energy  in the large-$Q^2$ region, and  for the longitudinally polarized virtual photon,   a QCD factorization theorem \cite{Collins:1996fb} dictates that the scattering amplitude can be written in terms of  the generalized parton distribution (GPD), the meson distribution amplitude (DA) and the perturbatively calculable hard part.  Even in cases where such a rigorous treatment is not  available, many phenomenologically successful models exist.   Generally speaking, light vector mesons probe the up and down quark contents of the proton, whereas in the Regge-regime $W\to \infty$, or for heavy vector mesons such as quarkonia ($J/\psi,\Upsilon,...$), the process is primarily sensitive to the gluonic content of the proton. 
 
The $\phi$-meson (a bound state of $s\bar{s}$) has a somewhat unique status in this context because its mass $m_\phi=1.02$ GeV, being roughly equal to the proton mass $m_N=0.94$ GeV, is neither light nor heavy. In the literature,  $\phi$-production is often discussed on a similar footing as quarkonium production. Namely, since the proton does not contain valence $s$-quarks,  the $\phi$-proton interaction proceeds mostly via gluon exchanges.  However, the proton contains a small but non-negligible fraction of strange sea quarks already on nonperturbative scales of $\sim 1 \, {\rm GeV}^2$. It is then a nontrivial question  whether the gluon exchanges, suppressed by the QCD coupling $\alpha_s$, always  dominate over the $s$-quark exchanges.  

In this paper, we take a fresh look at the lepto-production of $\phi$-mesons $e p \to e'p'\phi$ in the high-$Q^2$ region {\it near threshold}, namely when $W$ is very low and barely enough to produce a $\phi$-meson $W\gtrsim m_\phi+m_N= 1.96$ GeV.  Measurements in the threshold region have  been performed at LEPS  \cite{Mibe:2005er,Chang:2010dg,Hiraiwa:2017xcs} for photo-production $Q^2\approx 0$ and at the Jefferson laboratory (JLab) for both  photo- and lepto-productions  $4\gtrsim Q^2 \gtrsim 0$ \cite{Santoro:2008ai,Qian:2010rr,Dey:2014tfa}. One of the main motivations of these experiments was to study the nature of gluon  (or Pomeron) exchanges. Our motivation however is drastically different, so let us   quickly provide relevant background information: 

 It has been argued  
that the near-threshold photo-production of a heavy quarkonium (see 
\cite{Ali:2019lzf} for an  ongoing experiment) is sensitive to  the gluon gravitational form factors $\langle P'|T_g^{\mu\nu}|P\rangle$  where $T_g^{\mu\nu}$ is the gluon part of the QCD energy momentum tensor \cite{Frankfurt:2002ka,Hatta:2018ina,Hatta:2019lxo} (see also \cite{Xie:2019soz,Mamo:2019mka}).  In lepto-production at high-$Q^2$, this connection can be cleanly established by using the operator product expansion (OPE)  \cite{Boussarie:2020vmu}. The analysis \cite{Frankfurt:2002ka} of $J/\psi$  photo-production over a wide range of $W$ indicates that the $t$-dependence of these form factors (or that of the gluon GPDs at high energy $W$) is weaker  than what one would expect from the $Q^2$-dependence of the electromagnetic form factors. This suggests  that the gluon fields in the nucleon are more compact and localized than the charge distribution. 
Moreover, it has been demonstrated \cite{Hatta:2018ina,Hatta:2019lxo,Boussarie:2020vmu} that  the shape of the differential cross section $d\sigma/dt$ is sensitive to the gluon D-term $D_g(t)$ which, after Fourier transforming to the coordinate space, can be interpreted as an internal force (or `pressure') exerted by gluons  inside the proton \cite{Polyakov:2018zvc} (see however, \cite{Freese:2021czn}). The experimental determination of the parameter $D_g(t=0)$ is complementary to the ongoing effort \cite{Burkert:2018bqq,Kumericki:2019ddg,Dutrieux:2021nlz} to extract the $u,d$-quark contributions to the D-term $D_{u,d}$ from the present and future  deeply virtual Compton scattering (DVCS) experiments. Together they constitute the total D-term  
\beq
D(0)=D_u(0)+D_d(0) + D_s(0) + D_g(0)+\cdots. \label{mis}
\eeq
Only the sum is conserved (renormalization-group invariant) and represents a fundamental constant of the proton.  

The strangeness contribution to the D-term $D_s$ has received very little, if any, attention in the literature so far. However, a large-$N_c$ argument \cite{Goeke:2001tz} suggests the approximate flavor independence of the D-term $D_u\approx D_d$ (nontrivial because $u$-quarks are more abundant than $d$-quarks in the proton) which by extension implies that $D_s$ could be comparable to $D_{u,d}$ at least in the flavor SU(3) symmetric limit. It is thus a potentially important piece of the sum rule  (\ref{mis}) when the precision study of the D-term  becomes possible in future.  
The purpose of this paper is to  argue that the lepto-production of $\phi$-mesons near threshold is governed by the strangeness gravitational form factors including  $D_s$.   By doing so, we basically postulate that the $\phi$-meson  couples more strongly to the $s\bar{s}$ content of the proton than to the gluon content at least near the  threshold at high-$Q^2$, contrary to the prevailing view in the literature.
Also, by using the local version of the OPE following Ref.~\cite{Boussarie:2020vmu}, we do not work in the collinear factorization framework  \cite{Collins:1996fb} whose applicability to the threshold region does not seem likely. In the appendix, we briefly discuss the connection between the two approaches. Other  approaches to $\phi$-meson photo- or lepto-production near threshold can be found in  \cite{Laget:2000gj,Titov:2003bk,Strakovsky:2020uqs}.  

In Section II, we collect general formulas for the cross section of lepto-production. In Section III, we describe our model of the scattering amplitude inspired by the OPE and the strangeness gravitational form factors. The numerical results for the differential cross section $d\sigma/dt$ are presented in Section IV for the kinematics relevant to the JLab and the future electron-ion collider (EIC).


\section{Exclusive $\phi$-meson leptoproduction}

Consider exclusive $\phi$-meson production cross section in electron-proton scattering $ep\to e'\gamma^* p\to  e'p'\phi$. The $ep$ and $\gamma^*p$ center-of-mass energies are denoted by $s_{ep}=(\ell+P)^2$ and $W^2=(q+P)^2$, respectively. The outgoing $\phi$ has momentum $k^\mu$ with $k^2=m_\phi^2$.  The cross section can be written as 
\beq
\frac{d\sigma}{dWdQ^2}&=& \frac{\alpha_{em}^2}{4\pi} \frac{1}{16  (P\cdot \ell)^2 Q^4 P_{cm}}\int \frac{d\phi_\ell}{2\pi} L_{\mu\nu}\int dt \,\frac{1}{2}\sum_{spin}\langle P|J_{em}^\mu(-q)|P'\phi\rangle\langle P'\phi|J_{em}^\nu(q)|P\rangle ,
\eeq
where $J_{em}$ is the electromagnetic current operator, $m_N$ is the proton mass and  $P_{cm}$ is the proton momentum in the $\gamma^*p$ center-of-mass frame 
\beq
P_{cm}=\frac{\sqrt{W^4-2W^2(m_N^2-Q^2)+(m_N^2+Q^2)^2}}{2W}.
\eeq
For simplicity, we average over the outgoing lepton angle $\phi_\ell$, but its dependence can be restored  \cite{Arens:1996xw} if need arises.
We can then  write, in a frame where the virtual photon is in the $+z$ direction, 
\beq
\int \frac{d\phi_\ell}{2\pi} L^{\mu\nu}=\frac{2Q^2}{1-\epsilon} \left(\frac{1}{2}g^{\mu\nu}_\perp +\epsilon \varepsilon^\mu_L \varepsilon^\nu_L  \right) \label{diehl}
\eeq
where $g_\perp^{ij}=\delta^{ij}$ and 
\beq
\varepsilon_L^\mu(q) = \frac{1}{Q\sqrt{1+\gamma^2}} \left(q^\mu+ \frac{Q^2}{P\cdot q}P^\mu\right).
\eeq
is the polarization vector of the longitudinal virtual photon. 
$\epsilon$ is the longitudinal-to-transverse photon flux ratio 
\beq
\epsilon=\frac{1-y-\frac{y^2\gamma^2}{4}}{1-y+\frac{y^2}{2}+\frac{y^2\gamma^2}{4}}, \qquad \frac{1}{1-\epsilon}= \frac{2}{y^2}\frac{1-y+\frac{y^2}{2}+\frac{y^2\gamma^2}{4}}{1+\gamma^2}
\eeq 
where
\beq
y\equiv \frac{P\cdot q}{P\cdot \ell}, \qquad \gamma\equiv\frac{2x_Bm_N}{Q} = \frac{m_NQ}{P\cdot q}
\eeq
The parameter $\gamma$ accounts for the target mass correction which should be included in near-threshold production. 
The variable $y$ can be eliminated in favor of $s_{ep}$ using the relation
\beq
W^2=y(s_{ep}-m_N^2)+m_N^2-Q^2. \label{y}
\eeq
In the hadronic part, let us define  
\beq
\frac{1}{2}\sum_{spin} \langle P|J_{em}^\mu(-q)|P'\phi\rangle \langle P'\phi|J_{em}^\nu(q)|P\rangle \equiv -   ( {\cal M}_{\rho}^{\ \mu})^*{\cal M}^{\rho\nu},
\eeq
where the minus sign is from the vector meson polarization sum $\sum \varepsilon_V^\rho \varepsilon_V^{\rho'}=-g^{\rho\rho'}+\frac{k^\rho k^{\rho'}}{m_\phi^2}$. 
(The amplitude ${\cal M}$ satisfies the conditions $k_\rho {\cal M}^{\rho\mu}= {\cal M}^{\rho\mu}q_\mu=0$.) 
 Contracting with the lepton tensor (\ref{diehl}), we get 
\beq
 -\left(\frac{1}{2}g^{\mu\nu}_\perp +\epsilon \varepsilon^\mu_L \varepsilon^\nu_L  \right) {\cal M}^*_{\rho\mu}{\cal M}^\rho_{\ \nu}  &=&  -\left(\frac{1}{2}g^{\mu\nu}_\perp +\frac{\epsilon}{1+\gamma^2}\frac{Q^2}{(P\cdot q)^2}P^\mu P^\nu  \right)   {\cal M}^*_{\rho\mu}{\cal M}^\rho_{\ \nu}  \nn 
 &=& \left(\frac{1}{2}g^{\mu\nu} -\frac{\left(\frac{1}{2}+\epsilon\right)Q^2}{(1+\gamma^2)(P\cdot q)^2}P^\mu P^\nu\right) {\cal M}^*_{\rho\mu}{\cal M}^\rho_{\ \nu} ,
 \label{need}
\eeq
where we used 
\beq
g_\perp^{\mu\nu}=-g^{\mu\nu} + \varepsilon_L^\mu \varepsilon_L^\nu -\frac{q^\mu q^\nu}{Q^2} .
\eeq
We thus arrive at 
 \beq
 \frac{d\sigma}{dWdQ^2}=\frac{\alpha_{em}}{4\pi}\frac{W(W^2-m_N^2)}{(P\cdot \ell)^2 Q^2(1-\epsilon)} \int dt \frac{d\sigma}{dt}, \label{epcross}
 \eeq
 with\footnote{ In Ref.~\cite{Boussarie:2020vmu}, the authors calculated the spin-averaged cross section $g^{\mu\nu}  {\cal M}^*_{\rho\mu}{\cal M}^\rho_{\ \nu}$.  To directly compare with the  lepto-production data, one should rather use the formula (\ref{need}). }
 \beq
 \frac{d\sigma}{dt}= \frac{\alpha_{em}}{8(W^2-m_N^2)WP_{cm}} \left(\frac{1}{2}g^{\mu\nu} -\frac{\left(\frac{1}{2}+\epsilon\right)Q^2}{(1+\gamma^2)(P\cdot q)^2}P^\mu P^\nu\right) {\cal M}^*_{\rho\mu}{\cal M}^\rho_{\ \nu}  . \label{dt}
 \eeq
 Note that the $\gamma^*p$ cross section (\ref{dt}) depends on the $ep$ center-of-mass energy $s_{ep}$ because  $\epsilon$ depends on $y$, and $y$ depends on $s_{ep}$ as in (\ref{y}). 
Experimentalists at the JLab have measured  $\frac{d\sigma}{dW dQ^2}$ \cite{Santoro:2008ai}, and from the data they have reconstructed the differential cross section (\ref{dt}) and the total  cross section  $\sigma(W,Q^2)=\int dt\,  \frac{d\sigma}{dt}$. In the next section we present a model for the scattering amplitude ${\cal M}$.

\section{Description of the model}

Our model for the matrix element $\langle  P'\phi(k)|J_{em}^\nu(q)|P\rangle$ has been inspired by the recently developed new approach to the near-threshold production of  heavy quarkonia  such as $J/\psi$ and $\Upsilon$   \cite{Boussarie:2020vmu}. 
The main steps of  \cite{Boussarie:2020vmu} are summarized as follows.  One first relates the $J/\psi$ production amplitude $\langle  P'J/\psi(k)|J_{em}^\nu(q)|P\rangle$ to the correlation function of the charm current operator $J_c^\mu = \bar{c}\gamma^\mu c$  
\beq
\int dxdy e^{ik\cdot x -iq\cdot y}\langle P'|{\rm T}\{J_c^\mu(x)J_c^\nu (y)\} |P\rangle,
\eeq
in a slightly off-shell kinematics $k^2\neq m_{J/\psi}^2$. One then  
performs the operator product expansion (OPE)  in the regime   
$Q^2\gg |t|,m_{J/\psi}^2$ and picks up  gluon bilinear operators $\sim FF$. 
The off-forward matrix elements $\langle P'|FF|P\rangle$ are parameterized by the gluon gravitational form factors. These include  the gluon momentum fraction $A_g$ (the second moment of the gluon PDF), the gluon D-term $D_g$ and the gluon condensate (trace anomaly) $\langle F^{\mu\nu}F_{\mu\nu}\rangle$. 

When adapting this approach to $\phi$-production, we recognize a few important differences. First, we need to keep $s$-quark bilinear operators $\sim \bar{s}s$ rather than gluon bilinears. A quick way to estimate their relative importance in the present approach is to compare  the momentum fractions  $A_{s+\bar{s}}$ and $\frac{\alpha_s}{2\pi}A_g$. Taking $A_{s+\bar{s}}\approx 0.04$,  $A_g\approx 0.4$ \cite{Maguire:2017ypu} and $\alpha_s=0.2\sim 0.3$ for example,  we see that the  strange sea quarks are more important than gluons, though not by a large margin. In the appendix, we give a slightly improved argument and show that the $s$-quark contribution gets an additional factor $\sim 2$.  Second, the condition $Q^2\gg |t|$ is more difficult to satisfy.\footnote{This condition is needed in order to ensure that the large momentum $Q$ does not flow into the nucleon vertex so that one can perform the $JJ$ OPE. } For example, the momentum transfer at the threshold is 
\beq
|t_{th}|=\frac{m_N(m_V^2+Q^2)}{m_N+m_V}.
\eeq
When $Q^2\gg m_V^2$, $|t_{th}|$ is a larger fraction of $Q^2$ in $\phi$-production $m_V=m_\phi\approx m_N$ than in $J/\psi$-production $m_V=m_{J/\psi}\approx 3m_N$. As one goes away (but not too far away)  from the threshold, the region $Q^2\gg |t|$ does exist. In principle, our predictions are  limited to such regions, though in practice they can be smoothly extrapolated to  $|t|>Q^2$ as long as $|t|$ is not too small. 

We now perform the OPE. A simple calculation shows 
\beq
{\cal A}_s^{\mu\nu}&\equiv &i\int d^dr e^{ir\cdot q} \Bigl(\bar{s}(0)\gamma^\mu S(0,-r)\gamma^\nu s(-r) + \bar{s}(-r)\gamma^\nu S(-r,0)\gamma^\mu s(0) \Bigr) \nn 
&=& \frac{-4}{(q^2-m_s^2)^2}  \left( q^2g^{\mu\alpha}g^{\nu\beta}-g^{\mu\alpha}q^\nu q^\beta -g^{\nu\beta}q^\mu q^\alpha + g^{\mu\nu}q^\alpha q^\beta \right)T_{\alpha\beta}^s + \cdots \nn
&\to& \frac{-4}{(q^2-m_s^2)^2} 
(q\cdot k g^{\mu\alpha}g^{\nu\beta} -g^{\mu\alpha}k^\nu q^\beta -g^{\nu\beta} q^\mu k^\alpha + k^\alpha q^\beta g^{\mu\nu})T^s_{\alpha\beta}+\cdots,
\label{last}
\eeq
 where 
 \beq
 S(0,-r) = \int \frac{d^d q}{(2\pi)^d} e^{-iq\cdot r} \frac{i(\Slash q+m_s)}{q^2-m_s^2},
 \eeq
 is the $s$-quark propagator and 
 \beq
 T^s_{\alpha\beta}= i\bar{s}\gamma_{(\alpha} \overleftrightarrow{D}_{\beta)} s. \label{em}
 \eeq
 is the $s$-quark contribution to the energy momentum tensor  ($\overleftrightarrow{D}\equiv \frac{D-\overleftarrow{D}}{2}$).  We   neglect the $s$-quark mass $m_s$ whenever it appears in the numerator. However, we  keep it in the denominator  to regularize the divergence $\frac{1}{q^2-m_s^2}\sim \frac{1}{Q^2+m_s^2}$ just in case we may want to extrapolate our results to smaller $Q^2$ values in future applications. 
 In the last line of (\ref{last}), we have implemented minimal modifications to make ${\cal A}_s^{\mu\nu}$   transverse with respect to both $q^\nu$ and $k^\mu$ as required by gauge invariance. While this is {\it ad hoc} at the present level of discussion, we anticipate that  total derivative/higher twist operators  restore gauge invariance, similarly to what happens in deeply virtual Compton scattering (DVCS) \cite{Anikin:2000em}

Of course, even after restricting ourselves  to quark bilinears, there are  other operators that can contribute to (\ref{last}). Potentially important operators include the axial vector operator $\bar{s}\gamma_\mu \gamma_5 s$ and  the $s$-quark twist-two operators with higher spins. The former is related to the (small) $s$-quark helicity contribution $\Delta s$ to the proton spin in the forward limit. Its off-forward matrix element is basically unknown. The latter are discussed in the appendix where it is found that, unlike in the quarkonium case \cite{Boussarie:2020vmu}, the twist-two, higher-spin operators are not negligible for the present problem. To mimic their effect, we introduce an overall phenomenological factor of 2.5 in (\ref{last}).



To evaluate the matrix element $\langle P'|{\cal A}_s^{\mu\nu}|P\rangle$, we use the following parameterization of the gravitational form factors \cite{Kobzarev:1962wt,Ji:1996ek} 
\beq
\langle P'|T_s^{\alpha\beta}|P\rangle = \bar{u}(P')\left[A_s(t)\gamma^{(\alpha}\bar{P}^{\beta)}+B_s(t)\frac{\bar{P}^{(\alpha}i\sigma^{\beta)\lambda}\Delta_\lambda}{2m_N} + D_s(t)\frac{\Delta^\alpha \Delta^\beta-g^{\alpha\beta}\Delta^2}{4m_N} + \bar{C}_s (t)m_N g^{\alpha\beta}\right]u(P), \label{mom}
\eeq
where $\bar{P}=\frac{P+P'}{2}$, $\Delta^\mu=P'^\mu - P^\mu$ and $t=\Delta^2$. 
$D(t)/4$ is often denoted by $C(t)$ in the literature.  We neglect $B_s$ following the  empirical observation that the flavor-singlet   $B_{u+d}$ is unusually small (see, e.g., \cite{Hagler:2003jd}). 
We further set $\bar{C}_s=-\frac{1}{4}A_s$ assuming that the trace anomaly is insignificant in the strangeness sector. [However, this point may be improved as was done for gluons in \cite{Hatta:2019lxo,Boussarie:2020vmu}.] For the remaining form factors, we employ the dipole and tripole ansatze  suggested by the perturbative counting rules at large-$t$ \cite{Tanaka:2018wea,Tong:2021ctu}  
\beq
A_{s}(t)=\frac{A_{s}(0)}{(1-t/m_A^2)^2}, \qquad D_s(t) 
=  \frac{D_s(0)}{(1-t/m_D^2)^3}, \label{an}
\eeq
with $A_s(0)=A_{s+\bar{s}}=0.04$ as mentioned above. 
We use the same effective masses $m_A=1.13$ GeV and $m_D=0.76$ GeV as for  the gluon gravitational form factors from lattice QCD  \cite{Shanahan:2018nnv}. This is reasonable given that $s$-quarks in the nucleon are generated by the gluon splitting $g\to s\bar{s}$.\footnote{In \cite{Frankfurt:2002ka}  and more recently in  \cite{Wang:2021dis}, the authors fitted the $\phi$-meson photo- and lepto-production data using the form $d\sigma/dt \propto A^2(t)\propto 1/(1-t/m_A^2)^4$ and found that the mass parameter $m_A$ is consistent with the $J/\psi$ case. This partially supports our procedure.   }  The value $D_s(0)$ is our main object of interest, and is treated here as a free parameter. As mentioned in the introduction, even though $s$-quarks are much less abundant in the proton $A_s\ll A_{u,d}$, a large-$N_c$ argument suggests that the D-terms are `flavor blind' $D_s\sim D_u\approx D_d$ \cite{Goeke:2001tz}.\footnote{The prediction $|D_u+D_d|\gg |D_u-D_d|$ from large-$N_c$ QCD is supported by the lattice simulations \cite{Hagler:2003jd}. Interestingly, and in contrast, the $B$-form factor is dominantly a flavor nonsinglet quantity $|B_u-B_d|\gg |B_u+B_d|$ as already mentioned.
}  In the flavor SU(3) limit, and at asymptotically large scales, the relation ($C_F=\frac{N_c^2-1}{2N_c}=4/3$)
\beq
 D_s(0) \approx \frac{1}{4C_F} D_g(0),
\eeq
together with the recent lattice result for the gluon D-term $D_g(0)\approx -7.2$ \cite{Shanahan:2018nnv} gives $D_s(0)\approx -1.35$. We thus vary the parameter in the range  $0>D_s(0)>-1.35$, with a particular interest in the possibility that $|D_s|$ is of order unity.   Note that this makes $\phi$-production rather special, compared to light or heavy meson productions. If $A_s\ll A_{u,d,g}$ but $D_s\sim D_{u,d}$, the effect of the D-term will be particularly large in the strangeness sector.

Finally, the proportionality constant between ${\cal M}^{\mu\nu}$ and $\langle P'|{\cal A}_s^{\mu\nu}|P\rangle$ can be determined  similarly to the $J/\psi$ case (see Eqs.~(48,49) of \cite{Boussarie:2020vmu}). Using the 
$\phi$-meson  mass $m_\phi=$1.02 GeV and its leptonic decay width  $\Gamma_{e^+e^-}=1.27$ keV,  
we find 
\beq
\frac{e_s^4e^2m_\phi^4}{g_{\gamma\phi }^2 } = \frac{4\pi e_s^4 \alpha^2_{em}m_\phi}{3\Gamma_{e^+e^-}} \approx 2.21, \label{overall}
\eeq
where $e_s=-1/3$ and $g_{\gamma \phi}$ is the decay constant. 
There is actually an uncertainty of order unity in the overall normalization of the amplitude as mentioned in \cite{Boussarie:2020vmu} (apart from the factor of 2.5 mentioned above). 
This can be fixed by fitting to the total cross section data.  Then the shape of $d\sigma/dt$ is the prediction of our model.

\section{Numerical results and discussions}

We now present our numerical results  for the JLab kinematics with a 6 GeV electron beam ($\sqrt{s_{ep}}\approx 3.5$ GeV). The experimental data \cite{Santoro:2008ai} have been taken in the range
\beq
1.4<Q^2<3.8\, {\rm GeV}^2, \qquad |t-t_{min}|<3.6\, {\rm GeV}^2, \qquad 2<W<3\, {\rm GeV}, \label{mix}
\eeq
where $t_{min}$ is the kinematical lower limit of $t$ which depends on $Q^2$ and $W$.  Admittedly, even the maximal value $Q^2=3.8$ GeV$^2$ is not large from a perturbative QCD point of view.
However, (\ref{mix}) is the only kinematical window where the lepto-production data exist. Our model actually provides smooth curves for observables  in the above range of $Q^2$. 

\begin{figure}
  \includegraphics[width=0.6\linewidth]{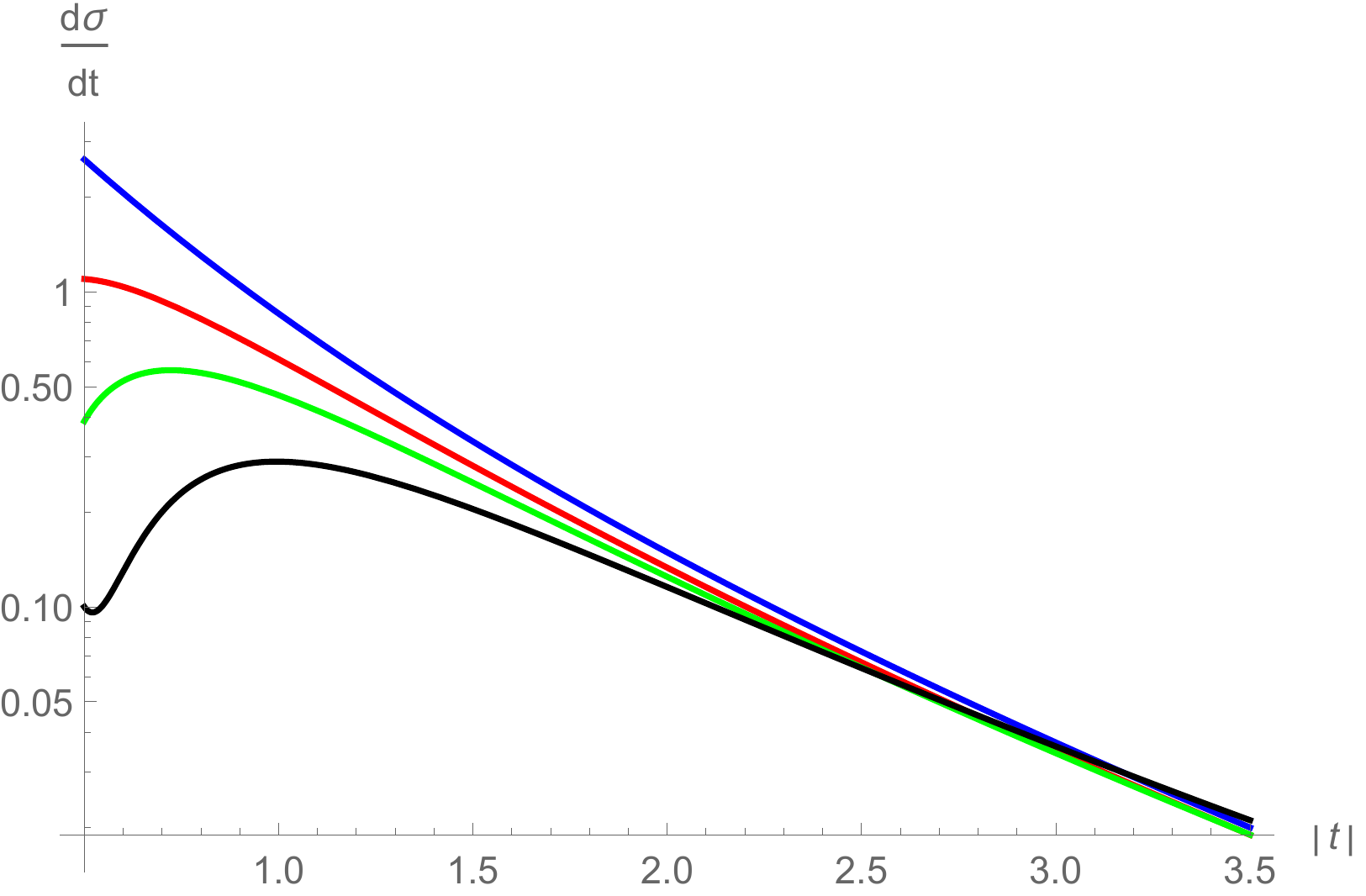}
\caption{Differential cross section $d\sigma/dt$ in units of nb/GeV$^2$ as a function of $|t|$ (in GeV$^2$). $W=2.5$ GeV, $Q^2=3.8$ GeV$^2$. The four curves correspond to $D_s=0,-0.4,-0.7,-1.3$ from top to bottom. 
}
\label{1}
\end{figure}

Fig.~\ref{1} shows  $d\sigma/dt$ at $W=2.5$ GeV, $Q^2=3.8$ GeV$^2$. 
We chose $m_s= 100$ MeV for the current $s$-quark mass.\footnote{ While this choice is natural in the present framework, it leads to a too steep rise of the cross section $\sigma(Q^2)$ as $Q^2$ is decreased towards $1$ GeV$^2$. Of course, our approach breaks down in this limit, but it is still possible to get a better  $Q^2$ behavior in the low-$Q^2$ region by switching to the constituent $s$-quark mass $m_s=m_\phi/2\approx 500$ MeV, or perhaps even $m_s\to m_\phi$ 
as in the vector meson dominance  (VMD) model.   }  The four curves correspond to different values of the $D$-term, $D_s=0,-0.4,-0.7,-1.3$ in descending order. 
We see that if the D-term is large enough, it causes a flattening or even a bump in the $|t|$-distribution in the small-$t$ region. This is due to the explicit factors of $\Delta^\mu$ ($t=\Delta^2$) multiplying $D_s$ in (\ref{mom}) which tend to shift the peak of the $t$-distribution to larger values.  Unfortunately, we cannot directly compare our result with the JLab data. The relevant plot, Fig.~18 of Ref.~\cite{Santoro:2008ai} is a mixture of data from different values of $Q^2$ in the range (\ref{mix}).  For a meaningful  comparison, $d\sigma/dt$ should be plotted for a fixed (large) value of $Q^2$, and there should be enough data points in the most interesting region  $|t|\lesssim 1$ GeV$^2$. By the same reason, we cannot adjust the overall normalization of the amplitude  mentioned at the end of the previous section. Incidentally, we note that in this kinematics the cross section is dominated by the contribution from the transversely polarized photon, namely, the part proportional to $g_\perp^{\mu\nu}$ in (\ref{need}).

\begin{figure}
  \includegraphics[width=0.6\linewidth]{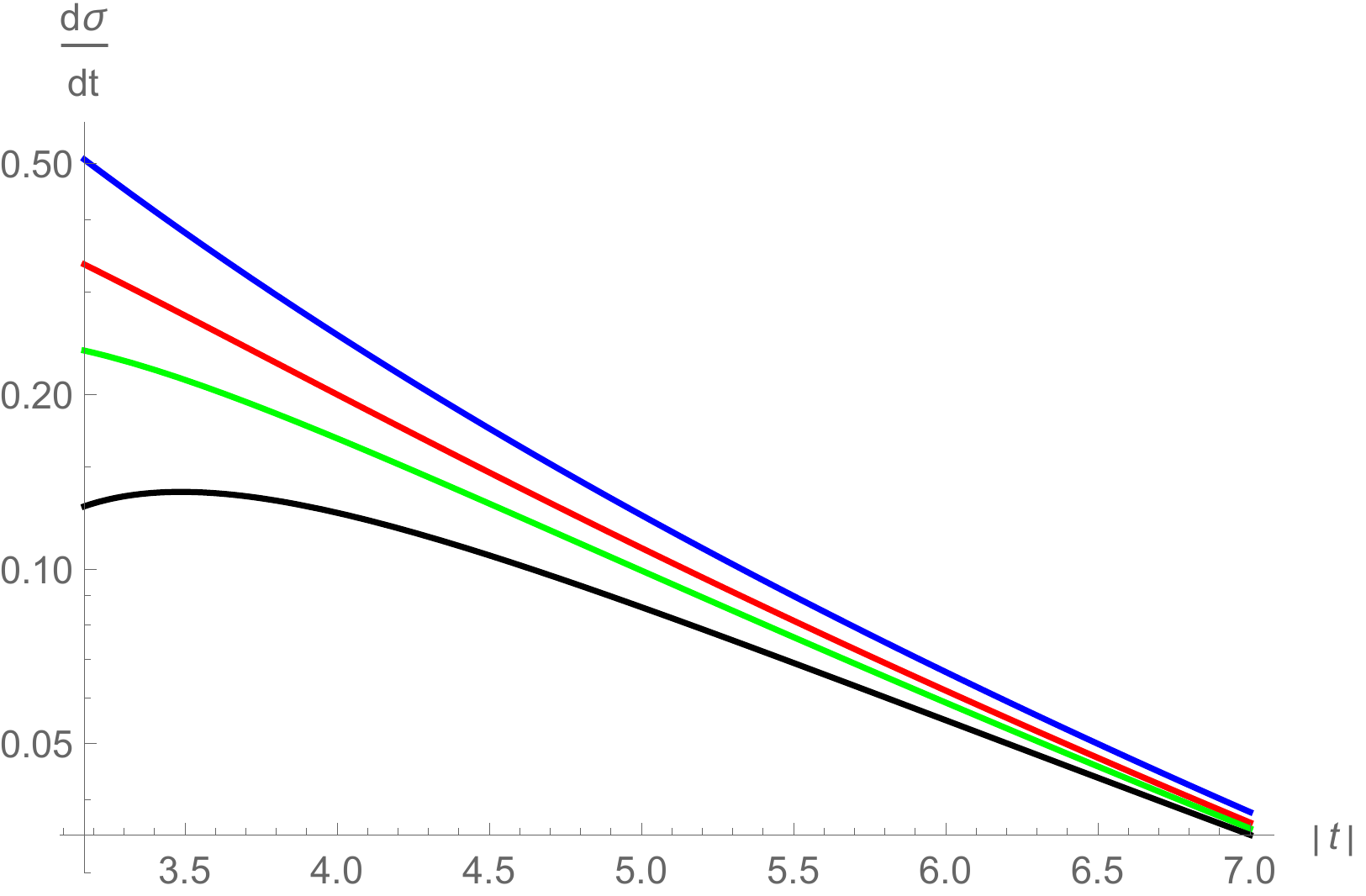}
\caption{Differential cross section $d\sigma/dt$ in units of pb/GeV$^2$ as a function of $|t|$ (in GeV$^2$). $W=2.5$ GeV, $Q^2=20$ GeV$^2$, $\sqrt{s_{ep}}=30$ GeV. The four curves correspond to $D_s=0,-0.4,-0.7,-1.3$ from top to bottom. 
}
\label{2}
\end{figure}

For illustration, in Fig.~\ref{2} we show the result with  $Q^2=20$ GeV$^2$ and  $W=2.5$ GeV, having in mind the kinematics of the Electron-Ion Colliders (EICs) in the U.S. \cite{Accardi:2012qut,Proceedings:2020eah} and in China \cite{Anderle:2021wcy}. We chose $\sqrt{s_{ep}}=30$ GeV for definiteness, but the dependence on $s_{ep}$ is very weak as it only enters the parameter $\epsilon$ in (\ref{dt}) and $\epsilon \approx 1$ in the present kinematics. (Of course the $ep$ cross section (\ref{epcross}) strongly depends on $s_{ep}$.) The contribution from the longitudinally polarized photon (the part proportional to $\epsilon \varepsilon_L^\mu \varepsilon_L^\nu$ in (\ref{need})) is now comparable to the  transverse part. 
Again the impact of the D-term is noticeable, but the bump has almost disappeared and  we only see a flattening of the curve in the extreme case $D_s=-1.3$. The reason is simple. The cross section schematically has the form 
\beq
\frac{d\sigma}{dt} \sim \frac{f(t)}{(1-t/m^2)^{a}},
\eeq
where $f(t)$ is a low-order polynomial in $t$ and $a=4,5,6$. (See (\ref{an}). The amplitude squared is a linear combination of $A_s^2(t),A_s(t)D_s(t)$ and $D_s^2(t)$.) The $t$-dependence of $f(t)$ comes from the D-term and gamma matrix traces involving nucleon spinors (\ref{mom}).   Clearly, $f(t)$ can affect the shape of $d\sigma/dt$ only when  $|t|< m^2\sim 1$ GeV$^2$. Beyond that, one simply has the power law $d\sigma/dt \sim 1/t^{c}$ with $c<4$.  In Fig.~\ref{1}, $|t_{min}|<1$ GeV, and this is why we see more interesting structures. As $Q^2$ gets larger, so does $|t_{min}|$ and the structure disappears.   \\

In conclusion, we have proposed a new model of $\phi$-meson lepto-production near threshold. In our model, the cross section is solely determined by the strangeness gravitational form factors, similarly to the $J/\psi$ case where it is determined by the gluon counterparts \cite{Boussarie:2020vmu}. Of particular interest is the value of $D_s$, the strangeness contribution to the proton D-term. While $D_s$ is ignored in most literature, an argument based on the large-$N_c$ QCD suggests that it may actually be comparable to  $D_{u,d}$ \cite{Goeke:2001tz}. If this is the case, we predict a flattening or possibly a bump in the $t$-distribution of $d\sigma/dt$ in the small-$t$ region. It is very interesting to test this scenario by  re-analyzing the JLab data \cite{Santoro:2008ai} or conducting new experiments focusing on the $|t|<1$ GeV$^2$ region.  


There are number of directions for improvement. As already mentioned, operators other than the energy momentum tensors should be included as much as possible, although this will unavoidably introduce more parameters in the model. We have argued in the appendix that the contribution from the twist-two, higher spin operators is small, but this needs to be checked. Also the renormalization group evolution of the form factors should be taken into account if in future one can measure this process over a broad range in $Q^2$ such as at the EICs in the U.S. and in China \cite{Accardi:2012qut,Anderle:2021wcy}.

\section*{Acknowledgments}
Y.~H. thanks the Yukawa Institute for Theoretical Physics for hospitality. 
This work is supported by the U.S. Department of Energy, Office of
Science, Office of Nuclear Physics, under contracts No. DE- SC0012704, DE-SC-0002145
and DE-FG02-93ER40771.
It is also supported in part by Laboratory Directed Research and Development (LDRD)
funds from Brookhaven Science Associates. 

\appendix
 \section{Connection to the GPD approach }

 In this appendix, we argue how our OPE approach is connected to the usual light-cone approach in terms of the generalized parton distribution (GPD), see, e.g., \cite{Goeke:2001tz} for a review. 
 Consider  doubly virtual Compton scattering (VVCS)  $p(P)\gamma^*(q) \to p(P')\gamma^*(q')$ or deeply virtual meson production (DVMP) $p(P)\gamma^*(q) \to p(P')V(q')$ and introduce the variables 
 \beq
 &&\bar{q}=\frac{q+q'}{2}, \qquad \bar{P}=\frac{P+P'}{2}, \qquad \Delta=P'-P=q-q', 
 \qquad \xi = \frac{-\bar{q}^2}{2\bar{P}\cdot \bar{q}} ,\qquad 
 \eta=\frac{-\Delta \cdot \bar{q}}{2\bar{P}\cdot \bar{q}}.
 \eeq
 $\eta$ is the skewness parameter and $\xi$ is the analog of the Bjorken variable $x_B=\frac{-q^2}{2P\cdot q}$ in DIS. 
In the regime of our interest $q'^2\sim m_\phi^2\ll -q^2=Q^2$, we can write 
 \beq
 \frac{\eta}{\xi} \approx  \frac{1}{1+\frac{\Delta^2}{2Q^2}}\approx 1 , \qquad  \eta \approx  \frac{x_B}{2-x_B\left(1 -\frac{\Delta^2}{Q^2}\right)} \approx \frac{x_B}{2-x_B},
 \eeq
 where the second approximation is valid when $Q^2\gg |t|=|\Delta^2|$.
  In DIS or DVCS, one takes the scaling limit   $-\bar{q}^2\to \infty$ and $2\bar{P}\cdot \bar{q}\to \infty$ (and implicitly $W^2=(P+q)^2\to \infty$) keeping the ratio  $1>\xi >0$ fixed. 
 However, near the threshold   where $W^2=(P+q)^2$ is constrained to be close to $(m_N+m_\phi)^2$, the scaling limit cannot be taken literally because   $x_B$ and $Q^2$ are no longer independent
\beq
x_B=\frac{Q^2}{2P\cdot q} =\frac{Q^2}{W^2+Q^2-m_N^2} \approx \frac{Q^2}{Q^2+2m_\phi m_N +m_\phi^2}.
\label{scaling}
\eeq
In the limit $Q^2\to \infty$, we have that 
\beq
x_B \approx \xi \approx \eta \approx 1.
\label{one}
\eeq
The partonic interpretation of scattering in this regime is rather peculiar.  Normally one works in a frame in which the incoming and outgoing protons are fast-moving 
\beq
\xi \approx \eta \approx \frac{P^+-P'^+}{P^++P'^+}. \label{ap}
\eeq
The incoming proton has the light-cone energy $P^+=(1+\eta)\bar{P}^+$, and it emits  two partons with momentum fractions
\beq
\frac{\eta+x}{1+\eta}, \quad \frac{\eta-x}{1+\eta}. 
\eeq
 When 
$\eta\approx \xi \approx 1$, $P^+\approx 2\bar{P}^+$ and the outgoing proton has vanishing light-cone energy $P'^+\approx 0$. Moreover, the condition $2P\cdot q\approx Q^2$ means $q^+\approx -P^+$ and $(P^++ q^+)q^-\sim m_N^2$. Therefore, the outgoing meson is not fast-moving in the minus direction $q^-\sim {\cal O}( m_N)$. Since the suppression of final state interactions due to large relative momenta is crucial for the proof of factorization \cite{Collins:1996fb}, we suspect that the standard approach based on GPDs is not applicable for near-threshold production, at least in its original form. 

 Nevertheless, we can make a rough connection to the present approach  as follows.    
When $\xi\approx 1$, the $s$-quark contribution to the Compton form factor may be Taylor expanded as
\beq
A_s(\xi,\eta)\sim \int^1_{-1} dx \left(\frac{1}{\xi-x-i\epsilon}-\frac{1}{\xi+x-i\epsilon}\right) H_s(x,\eta,t)\approx  2\int^1_{-1} dx (x+x^3+\cdots) H_s(x,1,t),
\eeq
where $H_s$ is the $s$-quark GPD. 
The lowest moment is proportional to the gravitational form factor $\langle P'|T_s^{++}|P\rangle$ that we keep, and higher moments give the form factors of the  twist-two higher spin operators $\langle x^n\rangle \sim \langle P'|\bar{s}\gamma^+(D^+)^n s|P\rangle$. 
To estimate the impact of the latter, let us substitute the asymptotic  form at large renormalization scales  \cite{Goeke:2001tz} 
\beq
H_s(x,\eta=1,t)\propto x(1-x^2).
\label{act}
\eeq
The above integral proportional to 
\beq
\int_{0}^1dx \frac{x^2(1-x^2)}{1-x^2}=\frac{1}{3}.
\eeq
If we only keep the first term in the Taylor expansion (corresponding to the energy momentum tensor), we get
\beq
\int_{0}^1dx x^2(1-x^2)=\frac{2}{15},
\eeq
that is,  40\% of the full result. This is in contrast to the $J/\psi$ case  \cite{Boussarie:2020vmu} where one has instead the gluon GPD 
\beq
\int dx \frac{H_g(x,\eta=1,t)}{1-x^2} \sim \int_0^1dx \frac{(1-x^2)^2}{1-x^2}=\frac{2}{3}.
\eeq
The first term in the Taylor expansion (corresponding to the gluon energy momentum tensor)
\beq
\int_0^1dx (1-x^2)^2=\frac{8}{15}.
\eeq
 accounts for 80\% of the total. The origin of this difference is easy to understand. The $s$-quark GPD $H_s$ vanishes at $x=0$ because the $s$ and $\bar{s}$ sea quarks are symmetric $H_{\bar{s}}(x)=-H_s(-x)=H_s(x)$. On the other hand, the gluon GPD $H_g$ is peaked at $x=0$ so higher moments in $x$ are numerically more  suppressed. 

We thus conclude that the twist-two, higher spin operators are not negligible in the $s$-quark case, although they are relatively innocuous in the gluon case. To cope with this, we introduce an overall factor $1/0.4=2.5$ in the leading order result (\ref{last}) as a model parameter. Note that the GPD $H_s$ contains a part related to the D-term \cite{Polyakov:1999gs}, so this factor is common to both the $A_s$ and $D_s$ form factors.

\end{document}